\documentclass[12pt, a4paper]{article}
\usepackage{amsmath, amssymb}
\usepackage{graphicx}
\usepackage{booktabs}
\usepackage{hyperref}
\usepackage{cleveref}
\usepackage{microtype}
\usepackage{fancyhdr}

\RequirePackage{tgtermes}
\RequirePackage{newtxtext}
\RequirePackage{newtxmath}
\RequirePackage{bm}
\RequirePackage{endnotes}

\usepackage{algorithm}

\usepackage{amsmath,amssymb}
\usepackage{graphicx}
\usepackage{array}
\usepackage{booktabs}
\usepackage{float}
\usepackage{caption}
\usepackage{natbib}
\usepackage{cite}
\usepackage{multirow}
\usepackage{comment}
\usepackage{enumerate}
\usepackage{enumitem}
\usepackage{makecell}
\usepackage{newfloat}
\usepackage{listings}

\usepackage{algorithm}
\usepackage{algpseudocode}
\usepackage{tikz}

\usepackage{natbib}
 \bibpunct[, ]{(}{)}{,}{a}{}{,}%
\usepackage{geometry}
\usepackage{array}
\usepackage{booktabs}
\usepackage{float}
\usepackage{caption}
\usepackage{cite}
\usepackage{amsmath}

\usepackage{graphicx}
\usepackage{enumerate}
\usepackage{array}

\usepackage{multirow}
\usepackage{comment}
\usepackage{mathtools}

\usepackage{amsfonts}
\usepackage{url}
\usepackage{amsthm}

\usepackage{algorithm}
\usepackage{algpseudocode}
\usepackage{tikz}

\usepackage{natbib}

\title{Learning to Collaborate: A Capability Vector-based Architecture for Adaptive Human-AI Decision Making}

\author{Renlong Jie\thanks{Dr. Renlong Jie is an Associate Professor at the School of Management, Northwestern Polytechnical University, Xi'an, China. Email: jierenlong@nwpu.edu.cn}}

\begin{document}

\maketitle

\begin{abstract}
Effective human–AI collaboration hinges on the ability to dynamically integrate the complementary strengths of human experts and AI models across diverse decision contexts. Context-aware weighted combination of human and AI outputs is a promising technique, which involves the optimization of combination weights based on capabilities of decision agents on a given task. However, existing approaches treat humans and AI as isolated entities, lacking a unified representation to model the heterogeneous capabilities of multiple decision agents. To address this gap, we propose a novel capability-aware architecture that models both human and AI decision-makers using learnable capability vectors. These vectors encode task-relevant competencies in a shared latent space and are used by a transformer-based weight generation module to produce instance-specific aggregation weights. Our framework supports flexible integration of confidence scores or one-hot decisions from a variable number of agents. We further introduce a learning-free baseline using optimized global weights for human–AI collaboration. Extensive experiments on image classification and hate speech detection tasks demonstrate that our approach outperforms state-of-the-art methods under various collaboration settings with both simulated and real human labels. The results highlight the robustness, scalability, and superior accuracy of our method, underscoring its potential for real-world applications. 
\end{abstract}

\section{Introduction}

With the rapid advancement of AI capabilities in recent years, human-AI collaborative systems have become an increasingly important topic of study. The emergence of large language models (LLMs) has enabled AI to perform a wide range of language-based tasks with remarkably high quality \citep{OpenAI2023GPT4}. However, AI systems still face limitations in areas requiring human-like nuanced judgment, real-world contextual understanding, value-based ethical considerations, or handling unstructured novel situations. Consequently, effectively integrating complementary human strengths with powerful AI capabilities presents a promising approach to enhance productivity and decision-making quality across diverse domains.

Existing studies have shown that these collaborative system can provide better decision quality compared with human-only or AI-only systems. For example, in skin cancer recognition, it is found that good quality AI-based support of clinical decision-making improves diagnostic accuracy over that of either AI or human doctors alone, and that the least experienced clinicians gain the most from AI-based support \citep{Tschandl2020Human}. In general image recognition, human-AI collaboration improves performance over human-only or AI-only systems through methods such as learning to defer \citep{Predict2018Madras, Hussein2020Consistent, verma2022calibrated, Krishnamurthy2023Enhancing} and probability merging \citep{kerrigan2021combining, Singh2023On}. For more complex tasks such as text generation, empirical evidence has shown that human-AI collaboration can improve both the efficiency and quality of writing \citep{Noy2023Experimental}. 

However, most existing studies on collaborative decision-making focus on limited scenarios and could fail to generalize across various decision agents with different capabilities. 
Meanwhile, traditional research often treats humans and AI models as independent entities in collaborative decision-making. Yet, advances in AI systems, such as modern computer vision models and large language models (LLMs), have endowed them with capabilities increasingly comparable to human decision agents \citep{he2015delving, redmon2016you, han2022survey, OpenAI2023GPT4, luo2025large}, narrowing the gap between human and AI decision-making capacities. In addition, the diversity between human decision makers is also an issue to be concerned \citep{lu2024mix}.
To uniformly model human and AI agent decision-making within a multi-agent framework, we introduce capability vectors. Analogous to recommendation system user/item embeddings \citep{zhang2019deep}, these vectors can be learned in an end-to-end manner. Building upon them, we develop a unified collaborative decision-making architecture supporting multiple human/AI agents via weighted merging of their output vectors. Experiments validate the effectiveness of both the architecture and the capability vectors.

The main contributions of our study include the following aspects: First, we introduce the learnable capability vectors to uniformly model the decision related capabilities of both different human experts and different AI models. In addition, we develop a learning method that can train the capability vectors as well as generating the decision weights of a set of decision agents for generating the final decision score.
Second, we propose a unified architecture for hybrid human-AI decision-making. This architecture leverages a weighted combination of output vectors from multiple decision agents and is capable of handling any multi-agent collaborative classification scenario, accommodating both one-hot and confidence-based outputs.
Third, we develop a learning-free baseline approach by transforming model outputs into one-hot vectors augmented with random noise. We then apply a probabilistic model to search for optimal global collaborative weights and predict the final accuracy.

The remainder of the paper is structured as follows. Section 2 reviews related work on human-AI collaboration mechanisms and capability modeling. Section 3 introduces our proposed methodology, including the architecture for weighted decision fusion, the learning mechanism for capability vectors, and an analysis of computational complexity. In section 4, we presents a learning-free baseline for multi-class classification using optimized global weights. Section 5 details extensive experiments on image classification and hate speech detection tasks, comparing our approach with state-of-the-art methods under various collaborative settings. In section 6, we provide an in-depth analysis of the model's behavior, including the effect of non-expertise capability levels and the learning dynamics of capability vectors. We further discuss the broader implications of our work and potential application scenarios in Section 7. Finally, we concludes the paper and suggests directions for future research in Section 8.

\section{Related Works}

\subsection{Collaboration Mechanisms}


Early work addressed AI deferring to single human experts, including two-stage rejection frameworks \citep{Predict2018Madras}, consistent surrogate losses for deferral \citep{Hussein2020Consistent}, and calibrated multiclass deferral \citep{verma2022calibrated}. Key findings show that well-calibrated AI confidence enhances collaboration \citep{vodrahalli2022uncalibrated}, optimizing individual AI accuracy does not necessarily optimize team performance \citep{Is2021Bansal}, and deferral notifications impact human-AI accuracy \citep{bondi2022role}. Complementary work includes taxonomies for analyzing complementary strengths \citep{rastogi2023taxonomy}, principled human teaching for deferral \citep{mozannar2022teaching}, and trust-driven reliance modeling \citep{Li2024modeling}.

Recent advances extend collaboration to multiple decision agents. \citet{kerrigan2021combining} integrated model probabilities with human class-level outputs, revealing dependencies on individual accuracies and confidence. \citet{Hemmer2022Forming} pioneered team formation strategies for human experts with instance allocation mechanisms. Ensemble techniques were further explored through human-AI label integration \citep{Singh2023On}, crowd wisdom principles \citep{da2020harnessing}, and model crowds \citep{he2022wisdom}. Methodological contributions include statistical ensemble assessment \citep{zou2024} and multi-expert deferral algorithms \citep{Mao2023Principled, mao2024two}.



\subsection{Capability Modeling}

There exists many metrics for evaluating human capabilities. 
For instance, the Intelligence Quotient (IQ) and Emotional Intelligence Quotient (EQ) are widely utilized to quantify the two major aspects of human intelligence \citep{1945The, Mayer2001Emotional}. Beyond unidimensional metrics, studies frequently adopt multidimensional frameworks like the Big Five personality model and the Myers-Briggs Type Indicator to represent personality, a key factor shaping decision-making styles \citep{Goldberg1990An, pittenger1993measuring, Myers1999MBTI}. 
Furthermore, language proficiency assessments like TOEFL, IELTS, and PETS, along with professional competency examinations such as the Bar Exam and CPA exam, evaluate human capabilities across different domains. Simultaneously, there is a growing body of research focused on assessing the capabilities of AI models. Task-specific models are typically evaluated using metrics relevant to their respective tasks, such as accuracy or recall for classification tasks, and BLEU or ROUGE scores for language generation tasks. For models designed to perform multiple tasks, their capabilities can be assessed through a combination of metrics derived from these various tasks.
Additionally, large language models (LLMs), which are capable of executing a wide range of language-based tasks akin to human performance, often require human evaluation for comprehensive assessment. Numerous benchmarks have been established for evaluating LLMs \citep{Chang2023A}. Some studies have even explored the application of IQ as a metric for evaluating AI models \citep{Liu2019How, Holzinger2019KANDINSKY, Kim2021Exploring}. 

As AI models increasingly exhibit capabilities that overlap with those of humans, it is of great value to build unified evaluation metrics for both entities. Such metrics should extend beyond singular indices like IQ and may take the form of learnable vectors, akin to token embeddings used in natural language processing, or user embeddings in recommender systems. 

\section{Methodology}

Human-AI collaborative decision-making has two primary objectives: enhancing the accuracy and quality of decisions, and reducing the associated decision-making costs. This study focuses on the former, specifically aiming to improve decision accuracy in classification tasks. To this end, it is essential to incorporate not only the final decision choices but also the confidence scores across candidate labels. In practice, however, human decision makers typically provide only their selected choices, whereas AI models can output predicted probabilities for all possible options. When these probabilities are well calibrated, they can serve as reliable confidence scores.

Existing learning-to-defer (L2D) methods generally assign decisions either to a human or an AI model, without effectively integrating their respective confidence scores. Moreover, the cost associated with human involvement remains dependent on the deferral rate. Recent research has demonstrated that combining human predictions with model probabilities can yield higher accuracy than relying on either humans or the model alone, with parameters of such a combination being efficiently estimable \citep{kerrigan2021combining}. At the same time, it has been observed that a naive integration of human and AI judgments may lead to suboptimal performance, underscoring the importance of selectively aggregating human inputs through intelligent strategies \citep{Singh2023On}. However, while existing methods often combine decisions based on human expertise modeled via confusion matrices, such representations exhibit notable limitations. They lack adaptability to complex multi-agent decision environments or classification tasks with large label spaces. Furthermore, confusion matrices offer a purely empirical and often superficial mapping of expert behavior, failing to capture underlying cognitive patterns. For instance, conditions such as color blindness, which significantly impact human performance in visual classification, cannot be adequately represented in a confusion matrix, yet they critically influence decision-making ability.

To avoid the shortcomings of existing methods, we apply instance-aware weighted combination of confidence scores and one-hot decision choices over decision options. The combination weights is determined by the capabilities of decision agents on each task instance. To better model the underlying capabilities of human or AI that may determine their decision performance in a collaborative system, we introduce learnable capability vectors with the following properties:
\begin{itemize}[leftmargin=2em]
\item A capability vector $\mathbf{C}\in \mathbb{R}^{1 \times d}$ is a 1-D vector with pre-defined length $d$.
\item The capability vector reflects all of a decision agent's capabilities related to a decision-making task category. It indicates that the subject's expected accuracy in a specific task within the category can be entirely determined by the capability vector and the task's attributes.
\item The similarity of capability vectors $c_i$ and $c_j$, measured by $\frac{\mathbf{c}_i \cdot \mathbf{c}_j}{||\mathbf{c}_i||\cdot ||\mathbf{c}_j||}$, indicates the correlation between the decision accuracies of different decision-making subjects in the same decision task. 
\end{itemize}
They can be applied in determining the optimal collaborative decision mechanism or decision weights for different decision tasks, and can be trained or pre-defined. This is similar to the user or item embedding in recommendation systems \citep{yu2016user, barkan2016item2vec, zhang2019deep, ghasemi2021user}, or word/token embedding in natural language processing \citep{kenton2019bert}. We assume that each candidate agent is associated with a capability vector, and the set of all these capability vectors forms a vocabulary. Overall, \textbf{our method is build upon the following three assumptions}: (a) There are many AI models or human experts with different capabilities, which can be represented by capability vectors with the same set of dimensions. (b) The optimal collaborative mechanism or decision weights for each agent is not fixed, but should be determined along with the other agents as well as the decision task. (c) The performance of collaborative decision making is determined by capability vectors of decision agents, the decision task and the collaborative decision mechanism.

\begin{figure*}[t]
\begin{center}
 \includegraphics[width=0.85\linewidth]{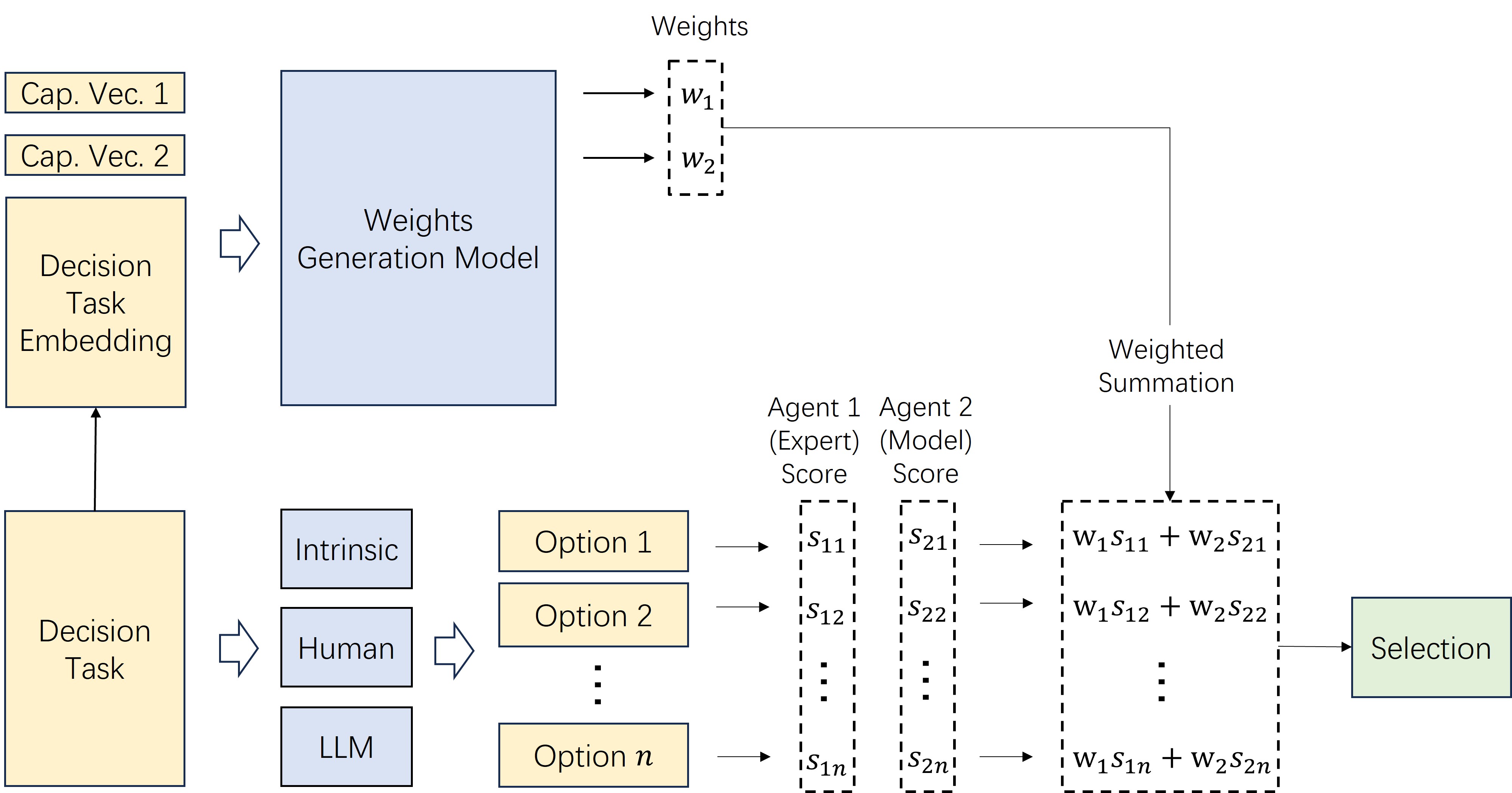}
 \caption{The diagram of the proposed architecture. The candidate decision options may be either pre-defined or generated by models or human experts. We employ a weighted aggregation of the scores assigned by each decision agent to each option in order to derive the final score for each choice. The weights assigned to each agent are determined by a weight generation model, which takes into account the capability vectors and the embedding of the decision task.} \label{Fig:1}
\end{center}
\end{figure*}

\subsection{Model architecture} \label{Sec:3.1}

We explore a generalized framework for human-AI decision-making.
Given a decision task associated with a decision context, we can get $n$ decision choices, which can be pre-defined or generated by a human-AI multi-agent system. 
Our objective is to identify the optimal choice among these options by leveraging a combination of human experts and AI models as decision agents. In this study, we propose to use a weighted linear combination of scores from all the decision agents to evaluate the goodness of each decision choice, and use that to rank the choices and make the final decision. The main architecture is shown in Figure ~\ref{Fig:1}. 

Assume that we have $m$ decision agents, each having a capability vector denoted as ${\mathbf{c}_1,...,\mathbf{c}_m}$. Then the weight generation model can give the weighting scores:
\begin{align}
w_1, w_2,...,w_m &= f(\mathbf{c}_1,...,\mathbf{c}_m;\mathbf{x};\mathbf{\theta}) \nonumber \\
\sum_i^m w_i &= 1 
\label{Eq:1}
\end{align}
where $f(.)$ is the weights generation model. Here we use a transformer encoder for this component \citep{vaswani2017attention}, while $\theta$ is the set of model parameter, and $x$ is the embedding of the decision task information. $w_1,...w_m$ are the decision weights for the decision agents. Usually, the size of a capability vector is different from that of the embedding of the input context. Thus, we add an extra linear transformation layer on the capability vectors or the task embeddings for aligning the dimensions $w_i=\mathbf{h}_{N_i}\mathbf{W}_{hw}$, $i=\{1,...,m\}$, where $\mathbf{W}_{hw}\in \mathbb{R}^{d\times 1}$ and $\mathbf{h}_{N_i} \in \mathbb{R}^{1\times d}$. Meanwhile, if the number of the decision agents $m$ is not fixed, a separation embedding between the capability embeddings and the decision context embeddings is needed. In addition, we apply a learnable positional embedding as in BERT to distinguish the positional information of each embeddings \citep{kenton2019bert}. 

In the side of decision agents, assume that there are $n$ possible decision outcomes, and each of them will be assessed by all of the agents and be assigned with $m$ scores. Therefore, we can get a score matrix $\mathbf{S}\in \mathbb{R}^{m\times n}$. Given the generated weights of the decision agents, the final scores of the candidate decision outcomes are given by: 
\begin{equation}
\mathbf{s}_f = \mathbf{w} \mathbf{S}
\label{Eq:2}
\end{equation}
where $\mathbf{w}=\{w_1,w_2,...,w_m\}\in \mathbb{R}^{1\times m}$, and $\mathbf{s}_f\in \mathbb{R}^{1\times n}$ is the vector of final scores of all the decision choices. We use the Softmax function to normalize the output, and then the weight of each category is given by $a_i = \frac{e^{s_{f_i}}}{\sum_i^m e^{s_{f_i}}}$,
where $i$ is the index of each category. Assume that the labels of the optimal decision among the candidate decision outcomes are given in the format of one-hot vector ${y^{(1)},...,y^{(N)}}$. Then we can use a cross-entropy loss for model training:
\begin{equation}
  L = -\frac{1}{N}\sum_{i=1}^N\sum_{j=1}^n y^{(i)}_k \log a^{(i)}_k
  \label{Eq:4}
\end{equation}
where $N$ is the number of examples in the labeled data batch. In general, the proposed architecture ensures that: (1) We apply a weighted combination of the decisions scores or decision selections from multiple decision agents with different capabilities, no matter humans or AI models, which include the case of probability combination as discussed in \citep{kerrigan2021combining}. (2) The decision weights depend on both the decision task and the capabilities of decision agents in a continuous space, which can be generalized to out-of-sample decision agents given the uniformly defined capability vectors. (3) Both the weight generation model and the capability vectors are learnable in an end-to-end manner given sufficient training data.


\subsection{Learning mechanism of capability vectors}

\begin{figure}[th]
\begin{center}
 \includegraphics[width=0.8\linewidth]{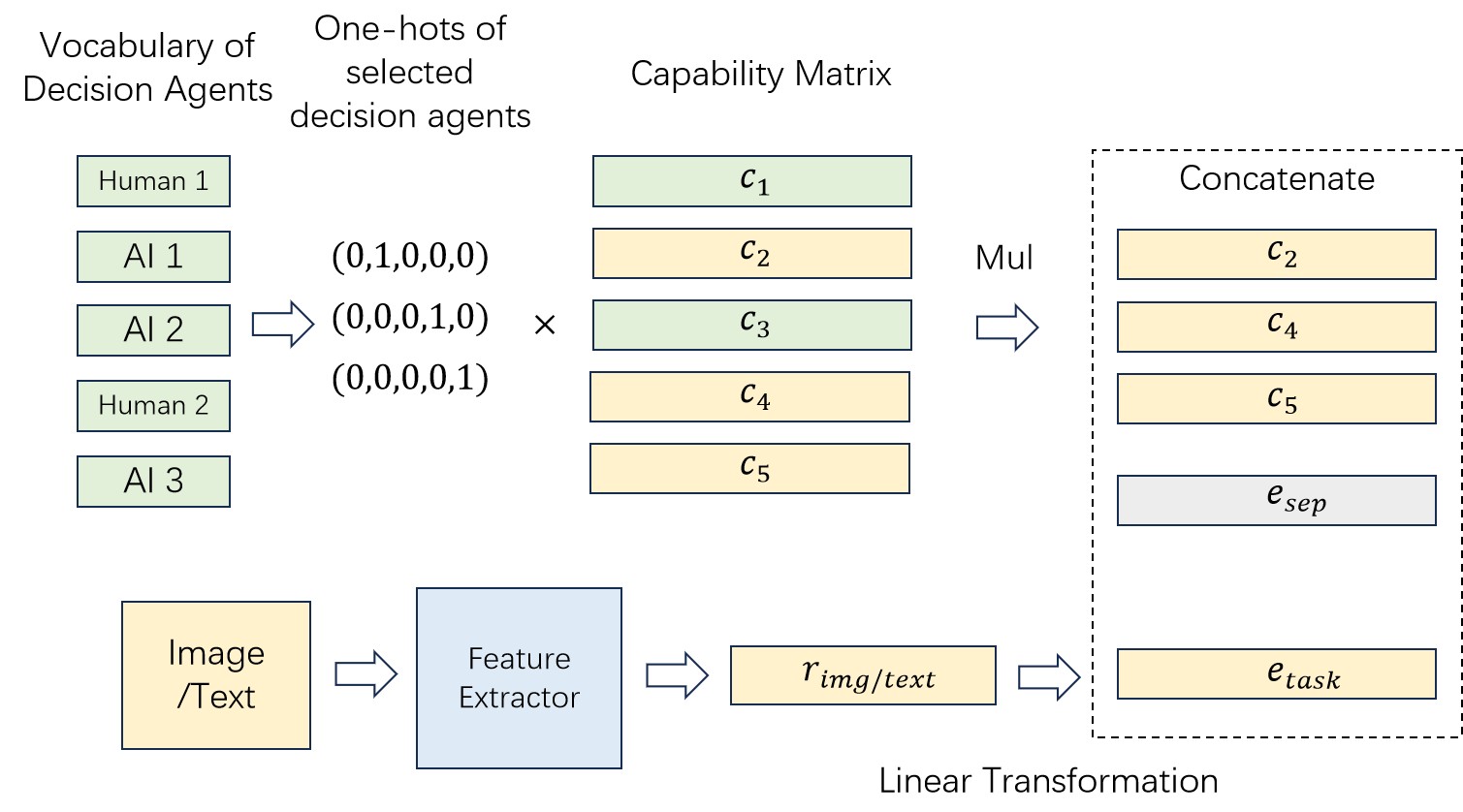}
 \caption{The diagram of the dimension alignment and the learning of capability vectors. At each iteration, we sample a subset of decision makers along with their corresponding one-hot vectors, to predict their respective weights. The final decision is derived from the weighted combination from all their scores. Furthermore, the capability matrix is subject to learning through back-propagation.} \label{Fig:2}
\end{center}
\end{figure}

The learning mechanism for capability vectors involves updating these vectors based on collaborative decision-making performance. As is shown in Figure~\ref{Fig:2}, each decision maker, whether a human expert or an AI model, is initially assigned a unique one-hot vector. This vector is then transformed into a capability vector through a linear transformation:
\begin{equation}
\mathbf{c}_i = \mathbf{e}_i \mathbf{W}_c
\end{equation}
Here, $\mathbf{e}_i$ is the one hot vector of $i$th expert, $\mathbf{W}_c \in \mathbb{R}^{d_{\text{vocab}} \times d_{\text{cap}}}$ is a learnable weight matrix, with $d_{\text{vocab}}$ representing the number of decision agents and $d_{\text{cap}}$ the dimensionality of the capability vectors. These capability vectors are fed into a weight generation model along with task embedding. Optimized capability vectors are crucial for determining optimal combination weights for candidate agents' outputs. If the capability vectors misrepresent the decision agent’s actual capabilities, the weight generation model cannot assign accurate weights, leading to reduced decision accuracy.
The final decision scores are computed by the weighted sum of the scores from all decision agents, using the generated weights. The capability vectors are updated through back-propagation using gradients derived from the final decision loss. The update rule for the capability matrix is:
\begin{equation}
\mathbf{W}_c \leftarrow \mathbf{W}_c - \eta \frac{\partial \mathcal{L}}{\partial \mathbf{W}_c}
\end{equation}
where $\eta$ is the learning rate and $\mathcal{L}$ represents the mini-batch loss function. Our learning mechanism employs random mini-batch sampling to train capability vectors, dynamically forming each batch by randomly sampling decision agents and tasks. This stochastic approach ensures the model generalizes well across different decision agents and tasks.

This mechanism resembles the training of token embeddings in language models and user embeddings in recommendation systems \citep{mikolov2013distributed, he2017neural}. However, capability vectors uniquely focus on decision-making performance across tasks. By refining capability vectors through context-based learning, we ensure they effectively capture the decision-making capabilities of both human experts and AI models.

\subsection{Computational Complexity Analysis}

The proposed dynamic task-aware collaboration framework exhibits the following computational complexity characteristics. The core operations are dominated by three components: (1) \textit{Capability vector learning} via linear transformation of one-hot decision agent identifiers Eq.~\eqref{Eq:4}, (2) \textit{Weight generation} through a Transformer encoder that processes capability vectors and task embeddings Eq.~\eqref{Eq:1}, and (3) \textit{Decision score aggregation} via weighted matrix combination Eq.~\eqref{Eq:2}. Key complexity parameters include: $m$ (active decision agents per task), $k$ is the length of task embedding (e.g. for image classification using ResNet output feature, $k=1$), the number of candidate decision choices $n$ , embedding dimension $d$ (512 by default), the number of transformer layers $L$ (6 by default), and the number of attention heads $h$ (8 by default). The quadratic $O(m^2)$ term in Transformer attention constitutes the primary scalability bottleneck for large $m$, though experimental results demonstrate feasibility up to $m=1000$ with standard hardware. The detailed comparison is shown in Table~\ref{tab:complexity}.


\begin{table*}[t]
\centering{\footnotesize
\caption{Computational complexity of framework components}
\label{tab:complexity}
\begin{tabular}{lccc}
\toprule
\textbf{Component} & \textbf{Operation} & \textbf{Time Complexity} & \textbf{Space Complexity} \\
\midrule
\multirow{2}{*}{Cap. Vec. Learning} 
    & One-hot to vector mapping & $O((m+k)d)$ & \multirow{2}{*}{$O(d_{\text{vocab}} d)$} \\
    & $\mathbf{c}_i = \mathbf{e}_i\mathbf{W}_c$ & per batch & \\
\midrule
\multirow{2}{*}{Weight Generation} 
    & Transformer encoder & $O\bigl(Lh\bigl((m+k)^2 d + (m+k) d^2\bigr)\bigr)$ & \multirow{2}{*}{$O\bigl(Lh\bigl((m+k)^2 + d^2\bigr)\bigr)$} \\
    & $f(\mathbf{c}_{1:(m+k)}, \mathbf{x}; \theta)$ & per instance & \\
\midrule
Decision Aggregation 
    & Weighted score fusion & $O((m+k)n)$ & $O((m+k)n)$ \\
    & $\mathbf{s}_f = \mathbf{w}\mathbf{S}$ & per instance & \\
\midrule
\multirow{2}{*}{\textbf{Overall System}} 
    & Training (end-to-end) & \makecell[l]{$O\bigl(Lh(m+k)^2 d + Lh(m+k) d^2$\\${}+(m+k)n\bigr)$} & \makecell[l]{$O\bigl(Lh(m+k)^2 + d^2$\\${}+(m+k)n\bigr)$} \\
    & Inference (per sample) & \makecell[l]{$O\bigl(Lh(m+k)^2 d + Lh(m+k) d^2$\\${}+(m+k)n\bigr)$} & $O\bigl(Lh(m+k)^2 + d^2\bigr)$ \\
\bottomrule
\end{tabular}}
\end{table*}

\subsection{Permutation invariance and positional encoding}


The proposed weight generation model (as described by Eq. ~\eqref{Eq:1}) demonstrates permutation invariance, meaning the order of decision agents or their capability vectors doesn't affect the final output architecture or performance. Mathematically, permuting indices in the capability vectors results in correspondingly permuted combination weights without altering overall performance:
\begin{align}
a(w_1,...w_m,\mathbf{s}_1,...,\mathbf{s}_m) \notag &=a(w_{\sigma(1)},...,w_{\sigma(1)}\\
&,\mathbf{s}_{\sigma(1)},...,\mathbf{s}_{\sigma(m)}) \\
a( f(\mathbf{c}_1,...,\mathbf{c}_m;\mathbf{x};\mathbf{\theta}),\mathbf{s}_1,...,\mathbf{s}_m) \notag &=a( f(\mathbf{c}_{\sigma(1)},...,\mathbf{c}_{\sigma(m)};\\
&\mathbf{x};\mathbf{\theta}),\mathbf{s}_{\sigma(1)},...,\mathbf{s}_{\sigma(m)})
\end{align}
where \(\sigma(1),..., \sigma(m)\) represents a permutation of the original indices \(1,...,m\). 
Given this permutation invariance, positional encoding is omitted in capability embeddings. Instead, a token type embedding differentiates capability embeddings from task embeddings, ensuring consistent model performance regardless of decision agent arrangement. This approach maintains the architecture's effectiveness while accommodating variable decision agent orders.

\section{A Learning-free Baseline for Multi-class Classification} \label{Sec:thm}

\subsection{Competence modeling of human experts.}

As discussed above, we use a vector to represent the competence of a human expert, which can be either learned or pre-defined, and should be mapped to the same dimension of the input embeddings. Here we discuss the modeling of real human expert's decision process, which means given a task instance, how is the expert's decision is generated.

In the case of classification problem, some existing works assume that a human expert can predict the correct label for a subset of classes, while making random selections for other classes \citep{Hussein2020Consistent}. However, in real world, an educated human expert does have the ability of doing some level of correct choice. For example, it is not likely that one classifies an animal as a vehicle, while sometime one may ``not sure'' if a photo is about one person or another. Therefore, we can assume that a human expert has one or multiple expertise categories, in which they can make close to 100\% correct decisions. For other categories, they can make correct decision with a limited probability $p$, and make random choice in other cases with probability $1-p$. 

\subsection{Expected decision accuracy of collaborative systems.}

We propose a learning-free baseline method by combining the model outputs with human predictions based on the competence model discussed above. Initially, we consider the scenario with one human expert and one AI model for collaborative decision making on classification tasks. Based on the above assumption for human experts, we can derive the weighted combination of model output and human expert's choice. In a straightforward case, we combine a one-hot vector representing the label selected by the human expert with the output vector generated by the AI model. 

\subsubsection{Building the combined output vector}

Let \( n \) denote the total number of categories. We define \( b_y \) and \( b_j \) as the output values corresponding to the true labeled class \( 1 \leq y \leq n \) and other classes \( 1 \leq j \leq n \) (\( j \neq y \)) in the output vector of the combined model, respectively. Let \( \alpha \) represent the combination weight of the model output, implying that the weight of the human expert's contribution is \( 1 - \alpha \). We consider two cases based on whether the true label \( y \) belongs to the expert's expertise set \( E \) or not, and calculate the combined output values for each index \( y \) and \( j \) in the combined output vector for both cases:
\begin{itemize}
\item Case 1: $y\in E$.
\begin{align}
    b_y &= \alpha a^1_y+ (1-\alpha) \nonumber \\
    b_j &= \alpha a^1_j \nonumber
\end{align}

\item Case 2: $y\notin E$.
\begin{align}
    b_y &= \alpha a^1_y+ (1-\alpha),\quad P = p + \frac{1}{n-K}(1-p) \nonumber\\
    b_y &= \alpha a^1_y,\quad P = \frac{n-K-1}{n-K}(1-p) \nonumber\\
    b_j &= \alpha a^1_j,\quad P = p + \frac{n-K-1}{n-K}(1-p) \nonumber\\
    b_j &= \alpha a^1_j+ (1-\alpha),\quad P = \frac{1}{n-K}(1-p) \nonumber
\end{align}

\end{itemize}

In Case 1, the true label \( y \) is situated within the categories of expertise \( E \) possessed by the human expert. In this scenario, the combined value at the index corresponding to the true label in the output vector is expressed as \( b_y = \alpha a^1_y + (1 - \alpha) \), where \( a^1_y \) denotes the model's output at index \( y \). If the human expert correctly predicts the label and assigns a value of 1 at index \( y \), the resulting combined value at this index becomes \( b_y \) as delineated above. Conversely, for any index \( j \neq y \), the model outputs a logit \( a^1_j \), while the human expert assigns a value of zero at index \( j \).

In Case 2, the true label lies outside the human expert's categories of expertise. The probability that the human expert assigns a value of 1 at the true label index \( y \) comprises two components: the probability \( p \) of making a correct decision despite the lack of expertise, and the probability of making a random yet correct guess. It is important to note that the expert can only make guesses among categories that fall outside their expertise set \( E \), as they are aware that the true label is not included in \( E \). Consequently, the probability of making a correct guess is given by \( \frac{1}{n-K}(1 - p) \), leading to the total probability that \( b_y \) receives a vote from the human expert being \( P_a = p + \frac{1}{n-K}(1 - p) \).

Following a similar rationale, we can ascertain the probabilities associated with other categories \( j \neq y \) receiving or not receiving a human expert's vote. For the sake of clarity in subsequent discussions, we define \( P_a = p + \frac{1}{n-K}(1 - p) \) and \( P_b = \frac{n-K-1}{n-K}(1 - p) \).

\subsubsection{Model output approximation}


Next, we model the output of the AI system within the collaborative framework. In classification models such as ResNet or BERT, the model generates a logits vector from which the category corresponding to the highest logit value is selected as the predicted label. We posit that the output logits vector of a classification model can be transformed into a one-hot vector representing the true label \( y \), augmented by random noise \( \delta_j \) added to all positions \( j \) in the vector. Consequently, the probability of obtaining a value \( \delta_j \) at any index \( j \neq y \) that exceeds $1+\delta_y$ corresponds to the error rate of the classification model.

To model the noise introduced to the one-hot representation, we employ a normal distribution \( \delta_j \sim N(0, \sigma^2) \), assuming that the noise terms are independent and identically distributed (i.i.d.). The process of deriving the predicted label from the model output under these assumptions is illustrated by:
\begin{equation*}
[a_1, \ldots, a_y, \ldots, a_n] \stackrel{\text{map}}{\rightarrow} [\delta_1, \ldots, 1 + \delta_y, \ldots, \delta_n] \notag 
\stackrel{\text{denoise}}{\rightarrow} [0, \ldots, 1, \ldots, 0]
\label{Eq:noise}
\end{equation*}
Here, the mapping function can be any order-preserving linear or nonlinear function shared across all instances. We assume the existence of an order-preserving function \( g(\cdot) \) that provides a sufficiently accurate approximation such that:
\[
g(a_1, \ldots, a_y, \ldots, a_n) \simeq [\delta_1, \ldots, 1 + \delta_y, \ldots, \delta_n]
\]
The function \( g \) can be estimated using various methods, including Maximum Likelihood Estimation, Bayesian Inference, or neural network approaches.

\subsubsection{Searching for the optimal combination weights}

Next, we focus on estimating the optimal values for the combination weights \( \alpha \) and \( 1 - \alpha \). We analyze two scenarios based on whether the expert provides a correct or incorrect vote concerning the ground-truth label. The weighted combined outputs for the label index \( y \) and the non-label index \( j \) are defined as follows:

\begin{itemize}
\item Case 1: The expert makes a correct vote on the ground-truth label.
\begin{align}
b_y &= \alpha(\delta_y + 1) + (1 - \alpha) = \alpha\delta_y + 1, \nonumber \\
b_j &= \alpha\delta_j \nonumber
\end{align}
In this case, the system arrives at the correct decision if and only if \( b_y > b_j \) for all \( j \neq y \). This condition can be expressed as \( \delta_y - \delta_j > -\frac{1}{\alpha} \).

\vspace{0.5em}
\item Case 2: The expert makes an incorrect vote, failing to vote for the ground-truth label.
\begin{align}
b_y &= \alpha(\delta_y + 1), \nonumber \\
b_j &= \alpha\delta_j + (1 - \alpha) \quad \text{or} \quad b_j = \alpha\delta_j \nonumber
\end{align}
In this case, if the \( j \)-th category receives a vote, the condition \( b_y > b_j \) translates to \( \delta_y - \delta_{j_1} > -\frac{1 - 2\alpha}{\alpha} \). Conversely, if the \( j \)-th category does not receive a vote, the condition simplifies to \( \delta_y - \delta_j > -1 \).
\end{itemize}
This analysis lays the groundwork for optimizing the combination weights \( \alpha \) and \( 1 - \alpha \) based on the performance of the expert's votes in relation to the ground-truth labels. We subsequently examine the overall probability that 
$b_y>b_j$ for any $j\neq y$ within the framework of the weighted combination strategy. Leveraging the relationship between joint and conditional probabilities, we can express the expected accuracy of the human-AI system as follows:
\begin{align}
P_1(b_y>b_j) =&P(y\in E, b_y>\max\{b_j\}_{j\neq y})+P(y\notin E, b_y>\max\{b_j\}_{j\neq y})\nonumber\\
=& P(y\in E)P(b_y>\max\{b_j\}_{j\neq y}|y\in E) + P(y\notin E)P(b_y>\max\{b_j\}_{j\neq y}|y\notin E)\nonumber\\
=& \frac{K}{n}P(\delta_y-\max\{\delta_j\}>-\frac{1}{\alpha}) + \frac{n-K}{n}(P(\delta_y-\delta_{{j}_1}>\frac{1-2\alpha}{\alpha}, \nonumber\\
&\delta_y-\max\{\delta_{j}\}>-1)P_b + P(\delta_y-\max\{\delta_j\}>-\frac{1}{\alpha})P_a)\nonumber\\
=& \frac{K}{n}\prod_{j\neq y} P(\delta_y-\delta_j>-\frac{1}{\alpha}) + \frac{n-K}{n}(P_b P(\delta_y-\delta_{{j}_1}>\frac{1-2\alpha}{\alpha}) \nonumber\\
&\prod_{j\neq y, j_1}P(\delta_y-\delta_{j}>-1) + P_a\prod_{j\neq y} P(\delta_y-\delta_j>-\frac{1}{\alpha})),
\label{Eq:11}
\end{align}
where $P_a = p + \frac{1}{n-K}(1-p)$ and $P_b= \frac{n-K-1}{n-K}(1-p)$ and \( K \) represents the size of the expert's expertise set \( |E| \). Under our assumptions, we model the difference \( \delta_y - \delta_j \) as following a normal distribution, specifically \( \delta_y - \delta_j \sim N(1, \sigma_D) \), with \( \sigma_D = \sqrt{2}\sigma \). Given that the true label is uniformly distributed, the probabilities associated with the expert's expertise set are \( P(y \in E) = \frac{K}{n} \) for falling within the expertise set (Case 1) and \( P(y \notin E) = \frac{n-K}{n} \) for falling outside of it (Case 2). In Case 2, the probabilities that the expert makes a correct or incorrect guess are denoted by \( P_a \) and \( P_b \), respectively. Furthermore, based on our modeling of the outputs from the AI system, the accuracy of the base AI model can be expressed as: $P_0(b_y>b_j) = P(\delta_y-\delta_j>-1)$.
where we assume that $\delta_y, \delta_j\sim N(0,\sigma)$, and the difference between two Gaussian random variables $\delta_y - \delta_j\sim N(0,2\sigma^2)$. Therefore, the accuracy of the AI model is given by:
\begin{align}
Acc&=P(b_y>\max\{b_j\}_{j\neq y}) \nonumber\\
&=(\int^{\infty}_{-1}\exp^{-\frac{x^2}{2\sigma_D^2}}dx)^{n-1} \nonumber\\
&= (\Phi(\frac{1}{\sigma_D}))^{n-1}
\end{align}
while $\sigma_D$ and $\sigma$ can be estimated with: $\sigma_D = 1/\Phi^{-1}((Acc)^{1/(n-1)})$.
For a model with a classification accuracy of $90\%$, the corresponding $\sigma_D=0.441$ and $\sigma=0.312$. Table~\ref{tab:sigma} shows the estimated values of $\sigma_D$ and $\sigma$ under a set of different model accuracies. 
\begin{table}[htbp]
    \caption{Estimated $\sigma_D$ and $\sigma$ under different model accuracies for 10-class classification.}
  \centering{\small
    \begin{tabular}{ccccccccccc}
    \toprule
    \textbf{Acc} & 0.5   & 0.55  & 0.6   & 0.65  & 0.7   & 0.75  & 0.8   & 0.85  & 0.9   & 0.95 \\
    \midrule
    $\sigma_D$ & 0.692  & 0.658  & 0.626  & 0.596  & 0.567  & 0.538  & 0.508  & 0.476  & 0.441  & 0.395  \\
    $\sigma$ & 0.489  & 0.465  & 0.443  & 0.422  & 0.401  & 0.380  & 0.359  & 0.337  & 0.312  & 0.279  \\
    \bottomrule
    \end{tabular}%
  \label{tab:sigma}}%
\end{table}%

\begin{figure*}[th]
\begin{center}
 \includegraphics[width=1.0\linewidth]{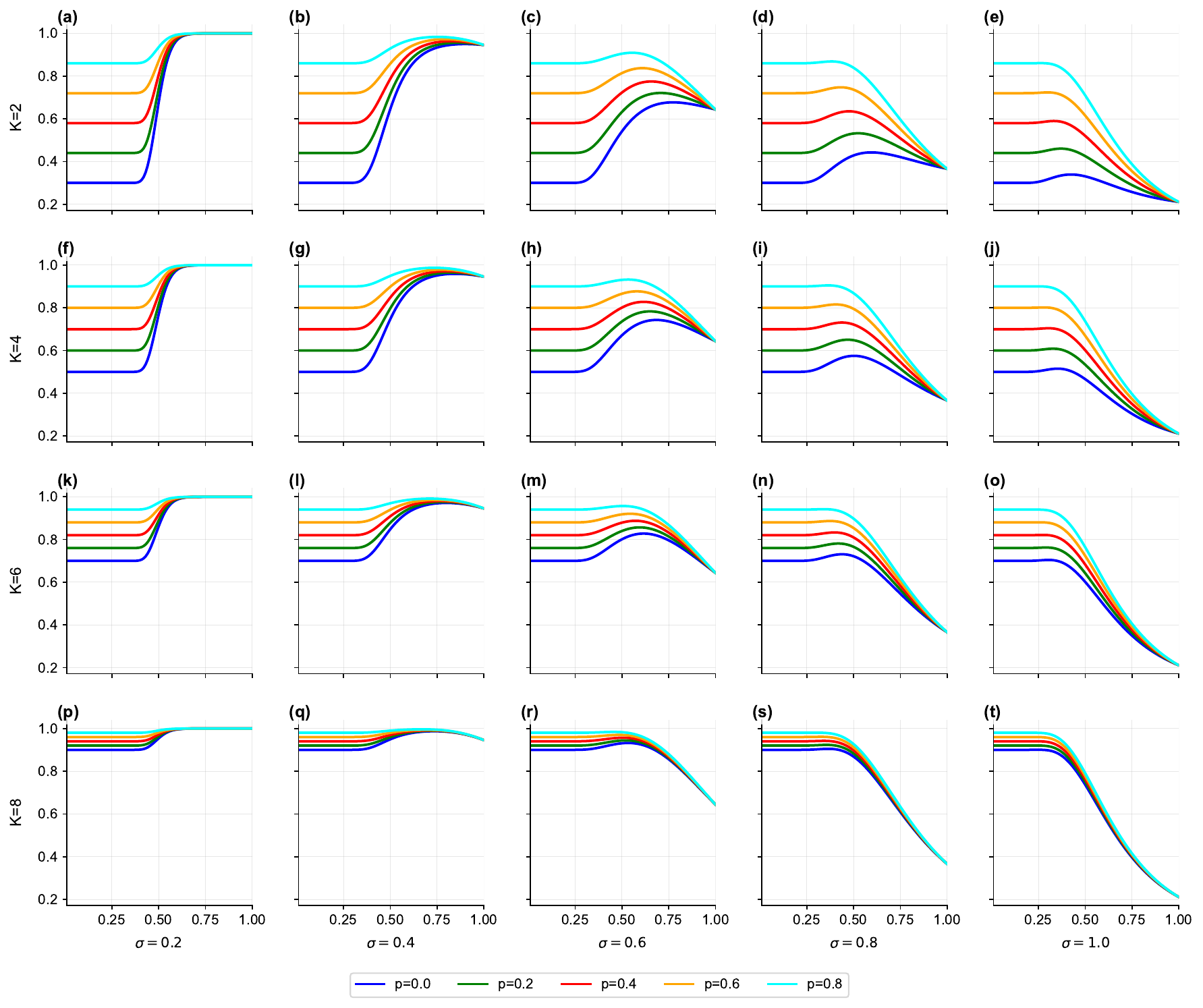}
 \caption{Estimated accurate probability under different collaborative settings. In each subplot, x-axis corresponds to different value of $\alpha$, and we utilize different colors for different values of $p$s.} \label{Fig:curve}
\end{center}
\end{figure*}

The condition for the human-AI system gives a better prediction than the base AI model is $P_1(b_y>b_j)> P_0(b_y>b_j)$, while the greater $P_1(b_y>b_j) - P_0(b_y>b_j)$, the higher advantage can be achieved by using combination collaborative strategy. Based on Eq.(\ref{Eq:11}), $P_1(b_y>b_j)>P_0(b_y>b_j)$ requires that:
\begin{align}
F(\alpha) =& \frac{K}{n}(\int^{\infty}_{-\frac{1}{\alpha}}\exp^{-\frac{x^2}{2\sigma_D^2}}dx)^{n-1} \nonumber\\
&+ \frac{n-K}{n}(P_b(\int^{\infty}_{\frac{1-2\alpha}{\alpha}}\exp^{-\frac{x^2}{2\sigma_D^2}}dx)(\int^{\infty}_{-1}\exp^{-\frac{x^2}{2\sigma_D^2}}dx)^{n-2}\nonumber\\ &+ P_a(\int^{\infty}_{-\frac{1}{\alpha}}\exp^{-\frac{x^2}{2\sigma_D^2}}dx)^{n-1})
- (\int^{\infty}_{-1}\exp^{-\frac{x^2}{2\sigma_D^2}}dx)^{n-1}\nonumber\\
&= \frac{K}{n}(\Phi(\frac{1}{\alpha\sigma_D}))^{n-1} + \frac{n-K}{n}(P_b(\Phi(\frac{2\alpha-1}{\alpha\sigma_D})\Phi(1/\sigma_D)^{n-2}\nonumber\\ &+ P_a(\Phi(\frac{1}{\alpha\sigma_D}))^{n-1})
- (\Phi(\frac{1}{\sigma_D}))^{n-1}\nonumber\\
>&0
\label{Eq:14}
\end{align}
where $\Phi(.)$ is the cumulative distribution of the standard normal density. To maximize $F(\alpha)$ w.r.t. $\alpha$, we can apply numerical methods such as grid search. We show the case with $n=10$ and varing combinations of $K$, $\sigma_D$ and $p$ as in Figure~\ref{Fig:curve}.
Our observations indicate that while different values of \( K \) do not significantly alter the shapes of the accuracy curves, the heights of these curves tend to increase with larger values of \( K \). In contrast, an increase in \( \sigma_D \), which signifies greater noise and consequently lower prediction accuracy from the AI model, results in a smaller optimal value of \( \alpha \). Notably, when both \( K \) and \( \sigma_D \) are large, the optimal \( \alpha \) may converge to zero, suggesting a reliance solely on the predictions of human experts. Furthermore, while variations in the non-expertise correct probability \( p \) may lead to minimal changes in the shapes of the accuracy curves, we observe that the accuracies at lower values of \( \alpha \) increase as \( p \) rises.

Overall, we establish a learning-free baseline by applying optimized global combination weights that integrate model outputs with human labels. This baseline serves as a benchmark to evaluate the effectiveness of instance-aware collaborative methods in multi-class classification tasks.

\section{Experiments}

\subsection{Datasets and Decision Tasks.}

We evaluate our method on four image classification datasets and one sentiment classification dataset for collaborative decision-making, including CIFAR10, CIFAR-10H, CIFAR100, GalaxyZoo, and Hatespeech. \textbf{CIFAR-10} \citep{Krizhevsky2009A} comprises 60,000 color images, each measuring 32x32 pixels. These images are categorized into 10 distinct classes, with each class containing 6,000 images. The dataset is divided into 50,000 training images and 10,000 test images. \textbf{CIFAR-100} consists of 60,000 32x32 color images spread across 100 classes, which are further organized into 20 super-classes \citep{Krizhevsky2009A}. Each class contains 600 images, with 5 classes per super-class. Each image is annotated with a ``fine'' label indicating its specific class and a ``coarse'' label representing its super-class. \textbf{Hatespeech} dataset consist of 24,783 tweets annotated as hate speech, oﬀensive language or neither \citep{davidson2017automated}. For our experiments with CIFAR-10 and CIFAR-100, we utilize the first 45,000 instances from the training set as the training data, reserving the remaining 5,000 instances for validation, while the original test set is employed as the test data. In the case of the Hatespeech dataset, we use the first 80\% of the data for training, the following 10\% for validation, and the rest 10\% for testing. In addition, CIFAR-10H and GalaxyZoo are datasets with real human labels. \textbf{CIFAR-10H} \citep{peterson2019human} is a human-annotated variant of the CIFAR-10 test set, where each of the 10,000 images has been labeled by multiple human annotators. \textbf{GalaxyZoo} is a crowd-sourced astronomical dataset for galaxy morphology classification \citep{lintott2008galaxy}. The detailed pre-processing procedure of these two datasets is given in Section~\ref{Sec:human_label}.

\subsection{Experimental settings}

We consider multiple scenarios of human-AI collaborative decision making, including collaborative decision by multiple human expert, collaborative decision by one model and one human expert, and collaborative decision by one model and multiple human experts.
We compare our method with traditional learning to defer methods, and check the effectiveness in improving the classification accuracy. We report the medium test accuracies of 5 trials using the checkpoints with the minimum validation loss in 200 epochs' of training. All experiments are run in Pytorch 2.3.0 with 2 Nvidia RTX 4090 GPUs.

\subsection{Collaborative decision by two human experts} \label{Sec:two_expert}

\subsubsection{Fixed decision makers}
At first, we consider a simple scenario. Assume we have two human experts. The first is familiar with animals, and the second is familiar with vehicles. However, we do not know the expertise of them before comparing their decision with the ground-truth labels. We assume that for CIFAR10 dataset, the \textbf{first human expert} can make correct classification on the images of birds, cats, horse, dog deer and frog, while \textbf{the second human expert} can make correct classification on the images of airplane, automobile, ship and truck. As there are only two decision makers, we can use two different capability vector to represent their decision capacity on image classification. We set $\mathbf{c}_1 = (0,1)$ and $\mathbf{c}_2 = (1,0)$, and map these two capability vectors to capability embedding vectors with linear transformations $c'_1 =  \mathbf{c}_1 \mathbf{W}_c$, $c'_2 =  \mathbf{c}_2 \mathbf{W}_c$, where $c'_1, c'2\in \mathbb{R}^{1\times d}$, $\mathbf{W}_c\in \mathbb{R}^{2\times d}$, and $d$ is the input embedding dimension. Meanwhile, we apply a pre-trained ResNet-18 to extract the feature of images \citep{he2016deep}, where the top-level representation has a length of 512. Thus we set the input embedding dimension to be 512.
We use a 4-layer 8-head transformer encoder with hidden size equal to 512. We use a transformer encoder as the weights generation model, and apply the segment embeddings to distinguish the positions of capability vectors and task embeddings. An Adam optimizer with a learning rate equals to 1e-5 is applied for optimization, while the batch size is set to be 64. We compare three settings of the feature extractor: a. \textbf{No task embedding}: We only apply capability vectors without using task embedding to determine the combination weights of decision makers, which results in the same combination weights for all task instance; b. \textbf{Fixed task embedding}: We use a finetuned Resnet-18 on CIFAR10 with fixed model parameters; c. \textbf{Full version}: We initialize the model with a Resnet-18 finetuned on CIFAR10, while the parameter can be further updated during the end-to-end training on collaborative decision making tasks. The results are given in Table~\ref{tab:1}. It shows that the proposed collaboration mechanism with tunable image representation extractor achieve the optimal classification accuracy. Given this finding, we use flexible feature extractors in the following experiments.
\begin{table}[htbp]
    \caption{Results of collaborative decision with two human experts on CIFAR 10 image classification task.}
  \centering{\small
    \begin{tabular}{cc}
    \toprule
    \textbf{Settings} & \textbf{Accuracy} \\
    \midrule
    No task embedding & 81.83 \\
    Fixed task embedding & 93.10 \\
    Full version & \textbf{97.82} \\
    \bottomrule
    \end{tabular}}%
  \label{tab:1}%
\end{table}%

\subsubsection{Random sampled decision makers} \label{Sec:6.3.2}

For changeable decision makers, the use of capability vector is essential as it is necessary to identify each decision maker's capability. We assume that there are $M$ candidate decision makers, while each time only a subset of them participate in collaborative decision making. Consider the multiple classification task with 10 categories as for CIFAR10. We further assume that each human expert can have expertise in one to ten categories, which means there will be $2^{10}=1024$ possible expert capabilities. Based on this assumption, we build \textbf{expert vocabularies with different sizes $M$} by sampling the expertise categories (with replacement) as a binary vector across all possible expert capabilities (i.e. from 1024 combinations). For each training or evaluation instance, we sample different 2 experts from the expert vocabulary for both training and testing. We consider three settings: (a) Use learnable capability matrix as is shown in Figure~\ref{Fig:2}; (b) Use fixed capability matrix; (c) Do not use capability vectors, just determine the two combination weights by the input image representations. We fixed the system random seed to 42 and compare different settings with the shared expert vocabulary. We apply an Adam optimizer with a learning rate of 1e-5 and the batch size equals to 64. The results are shown in Table~\ref{tab:cap_diff_exp}.

\begin{table}[htbp]
\caption{The effect of learnable capability vectors with different number of experts $M$ in the vocabulary. }
  \centering{\small
    \begin{tabular}{ccccccc}
    \toprule
    Size  & Settings & 10    & 30    & 100   & 300   & 1000 \\
    \midrule
    \multirow{3}[2]{*}{Small} & Learnable & \textbf{71.34} & \textbf{70.12} & \textbf{71.18} & \textbf{70.89} & \textbf{71.83} \\
          & Fixed & 71.18 & 69.72 & 70.64 & 69.16 & 69.25 \\
          & None  & 66.01 & 66.08 & 67.52 & 67.7  & 68.21 \\
    \midrule
    \multirow{3}[2]{*}{Large} & Learnable & 73.04 & \textbf{73.36} & \textbf{74.91} & \textbf{74.53} & \textbf{73.16} \\
          & Fixed & \textbf{73.15} & 73.32 & 74.84 & 74.22 & 71.39 \\
          & None  & 65.85 & 66.07 & 67.66 & 67.93 & 68.39 \\
    \bottomrule
    \end{tabular}}%
  \label{tab:cap_diff_exp}%
\end{table}%

We notice that given the same model size, as the vocabulary size of decision makers becomes larger, the advantage of using learnable capability vectors becomes larger. Notice that the weight of capability matrix only takes less than 1\% extra parameters of the weight generation model. This finding can be applied in building collaborative annotation systems with large number of candidate annotators. Moreover, the learning behavior of the capability vectors is investigated in Section~\ref{Sec:7.2}. 

\subsection{Collaborative decision by one model and one human-expert.}

To investigate the performance of our method on human-AI collaborative decision making, we initially focus on a scenario involving a single AI model and a human expert. Utilizing the strategy outlined in Section~\ref{Sec:two_expert}, we employ two one-hot vectors to differentiate between the AI model and the human expert. Subsequently, we apply a linear transformation to convert these vectors into their corresponding learned capability vectors, ensuring that they match the dimensionality of the task representation embedding.
Our collaborative decision-making framework is then implemented and assessed across three classification datasets: CIFAR-10, CIFAR-100, and Hatespeech.


\subsubsection{Results on CIFAR10} \label{Sec:cifar10}


In the first experiment, we follow \citep{Hussein2020Consistent} to simulate multiple synthetic experts of varying competence: if the image belongs to the ﬁrst $k$ classes with $k=1,2,...,10$,  the expert predicts the correct label of the image, otherwise the expert predicts uniformly over all classes. We use a tunable resnet-18 pre-trained on CIFAR-10 as the feature extractor, and a 28-layer Wide-Resnet same with that in \citep{Hussein2020Consistent} is applied as the AI classifier to be collaborative with human expert. 
To ensure that the AI classifier is the same for our method and the baseline method, we reproduce the baseline method in \citep{Hussein2020Consistent} with 200 epochs of training, and then use the weights of the AI classifier to initialize the classifier of our proposed model. For training the whole architecture, we apply an Adam optimizer with a learning rate of 1e-5 and a batch size of 64, with 200 epochs of training. We take different levels of expert's competence $k$, and compare the performance of our method with a set of baseline methods including Global (the proposed global weighted output merging methods in Section~\ref{Sec:thm}), Consistent \citep{Hussein2020Consistent}, Confidence \citep{raghu2019algorithmic}, OracleReject and MPZ18 \citep{Predict2018Madras}. The results are given in Table ~\ref{tab:2}. We can learn that our proposed architecture provide a consistent improvement from the baseline methods. In addition, there is no extra hyper-parameter involved in the loss function to be optimized. We also notice that if we extend the training epochs to 500 for the baseline method in \citep{Hussein2020Consistent}, gain of the system accuracy is not significant.

\begin{table}[htbp]
\caption{Comparison of system accuracy between our method and baselines with varying expert competence on CIFAR10. Among the baseline methods, for Consistent method, we reproduce the experiments by ourself. For other methods, we directly apply the reported results in \citep{Hussein2020Consistent}.}
  \centering{\small
    \begin{tabular}{ccccccccccc}
    \toprule
    k     & 1     & 2     & 3     & 4     & 5     & 6     & 7     & 8     & 9     & 10 \\
    \midrule
    Ours  & 95.53 & 95.85 & 96.16 & 97.21 & 97.66 & 98.47 & 98.96 & 99.25 & 99.66 & 100 \\
    Global & 94.68 & 94.99 & 95.40 & 95.89 & 96.46 & 97.11 & 97.85 & 98.71 & 100.0 & 100 \\
    Consistent & 94.63 & 95.26 & 95.53 & 95.89 & 96.31 & 98.08 & 98.71 & 99.14 & 99.42 & 100 \\
    Confidence & 90.47 & 90.56 & 90.71 & 91.41 & 92.52 & 94.15 & 95.5  & 97.35 & 98.05 & 100 \\
    OracleReject & 89.54 & 89.51 & 89.48 & 90.75 & 90.64 & 93.25 & 95.28 & 96.52 & 98.16 & 100 \\
    MPZ18 & 90.4  & 90.4  & 90.4  & 90.4  & 90.4  & 90.4  & 90.4  & 94.48 & 95.09 & 100 \\
    \bottomrule
    \end{tabular}}%
  \label{tab:2}%
\end{table}%

To further investigate the effect of human expert with relatively larger non-expertise capability level, we fix $k=5$ and change the non-expertise capability level by using different probability $p$. Using the same optimizer and hyper-parameter setting, the results are given in Figure~\ref{Fig:cap_level} (a). We observe that our method achieve higher accuracy compared with the baseline models. Again, the proposed global weighted merging baseline achieves competitive performances compared with instance aware methods, but still outperformed by the proposed instance aware weighted merging architecture. In addition, we observe that as the non-expertise capability level increase, the baseline L2D method in \citep{Hussein2020Consistent} has an accuracy drop when $p$ lies between 0.5 and 0.9. Meanwhile, the our proposed method, the system accuracy generally increases with the increase of non-expertise capability level. One possible explanation is that as the non-expertise capability level get close to the model accuracy, the baseline method may have instability in learning or be sensitive to hyper-parameter settings, which can be better resolved by merging the output of the model with human expert's decision.

\subsubsection{Results on CIFAR100.} \label{Sec:cifar100}

\begin{figure}[t]
\begin{center}
 \includegraphics[width=1.0\linewidth]{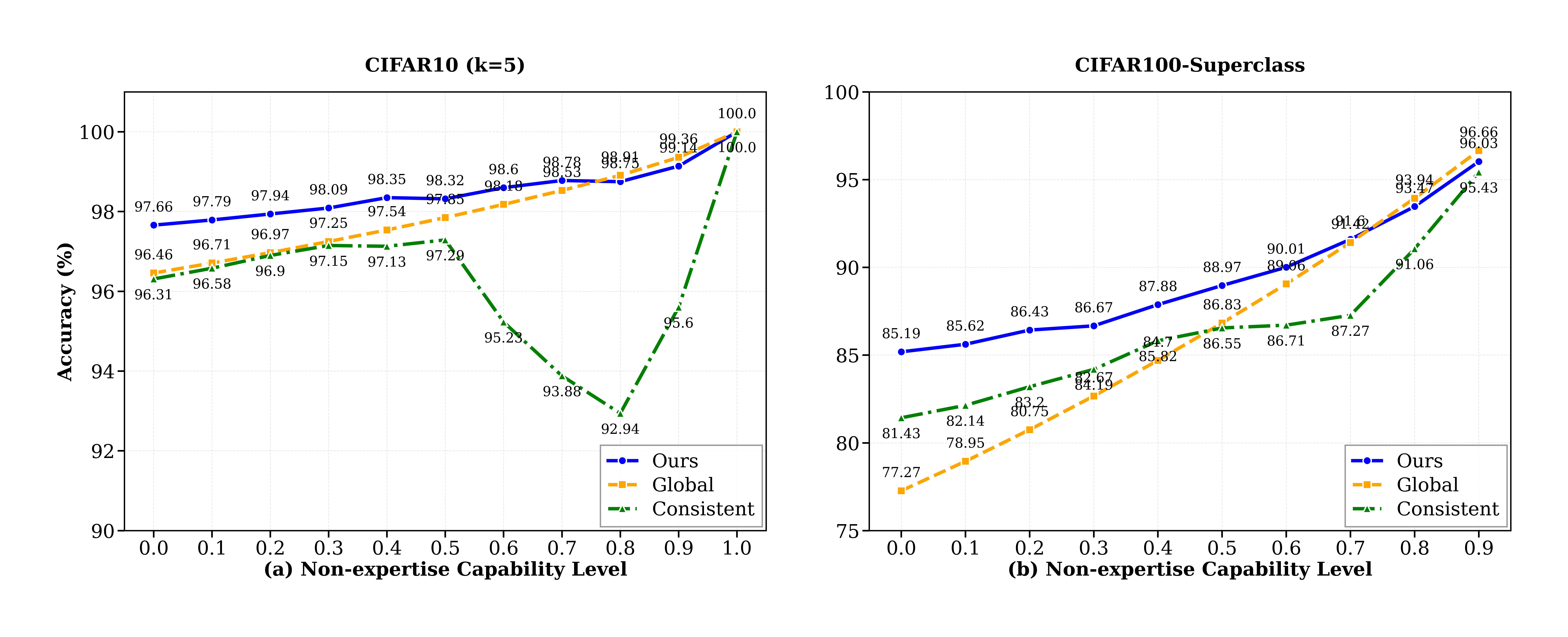}
 \caption{Diagram of accuracy under different non-expertise capability levels.} 
 \label{Fig:cap_level}
\end{center}
\end{figure}

We further conduct an experiment on CIFAR 100, in which there are 100 classes and 20 super-classes of images. Instead of predicting the labels of fine-grained classes, we predict the labels of super-classes. We train a Wide-ResNet-28 on this classification task and get an accuracy of 73.33\%, and utilize the pretrained model to initialize the AI classifier in both our proposed architecture and the baseline learning to defer method. We then compare our method with two alternative approaches:(a) Global, weighted output merging approach proposed in Section~\ref{Sec:thm}; (b) Consistent learning to defer estimator \citep{Hussein2020Consistent}, where each time the decision is made by only one expert or AI model. We tune the hyper-parameter of learning rates for both settings with a set of $\{1e^{-3}, 3e^{-4}, 1e^{-4}, 3e^{-5}, 1e^{-5}, 3e^{-6}, 1e^{-6}\}$. The results of varying expert competence is given in Table~\ref{tab:cifar100_ve}.
\begin{table}[htbp]
\caption{Experiment on CIFAR100-superclass with varying expert competence.}
  \centering{\small
    \begin{tabular}{cccccccccccc}
    \toprule
          & 0     & 2     & 4     & 6     & 8     & 10    & 12    & 14    & 16    & 18    & 20 \\
    \midrule
    Ours  & 73.33 & 78.95 & 79.68 & 81.2  & 82.99 & 85.19 & 87.01 & 91.06 & 94.39 & 97.43 & 100 \\
    Global  & 73.33 & 74.30 & 76.17 & 78.45  & 81.02 & 83.81 & 86.83 & 90.06 & 93.56 & 97.51 & 100 \\
    Consistent & 73.33 & 76.99 & 78.61 & 79.65 & 82.25 & 81.43 & 86.06 & 83.1  & 89.37 & 94.36 & 100 \\
    \bottomrule
    \end{tabular}} 
  \label{tab:cifar100_ve}%
\end{table}%
It is shown that with the same initialized AI classifier and the experts' competences, our method outperforms the baseline L2D method significantly in the cases of effective human-AI collaboration. Global weighted merging method achieve very competitive results as $k$ gets larger. Similary as in Section~\ref{Sec:cifar10}, we further investigate the effective of varying non-expertise capability levels with expert competence level $k=5$.
The results are shown in Figure~\ref{Fig:cap_level} (b). We observe that our method has a greatest advantage over the baseline around $p=0.7$, and for other non-expertise capability levels, it is possible that the advantage is smaller than the advantage at $p=0.0$.

\subsubsection{Results on HateSpeech} \label{Sec:hatespeech}

We also conducted experiments on Hate speech dataset, in which we apply the proposed architecture and baseline methods to determine if an instance is hatespeech, offensive language or neither. We follow \citep{Hussein2020Consistent} to create the synthetic experts for this task, which is: (a) If the tweet is in AAE then with probability $p$
the expert predict the correct label and otherwise predict uniformly at random. (b) If the
tweet is not in AAE, the expert predict the correct label with probability $q$ and otherwise predict uniformly at random. Also, we consider three different expert probabilities $p$ and $q$, corresponding to a fair expert ($p=q=0.9$), a biased expert towards AAE ($p=0.75, q=0.9$), and a biased expert towards non AAE tweets ($p=0.9, q=0.75$). We use a pre-trained transformer encoder on Hatespeech to initialize our model as well as the classifier in the baseline model. We use a Adam optimizer with a learning rate of 1e-6 and a batch size of 32. The results are shown in Table~\ref{tab:hatespeech}. The collaborative method largely outperforms the model-only and human-only methods. Moreover, compared with the method of consistent estimator given in \citep{Hussein2020Consistent}, our method achieves significant improvements for non-AAE biased and fair setting. 

\begin{table}[htbp]
 \caption{Results for our method and baselines on the hate speech detection task.}
  \centering{\small
    \begin{tabular}{cccc}
    \toprule
    p,q   & 0.75,0.9 & 0.9,0.75 & 0.9,0.9 \\
    \midrule
    Ours  & 94.96 & \textbf{94.62} & \textbf{95.88} \\
    Consistent & \textbf{95.14} & 93.68 & 95.63 \\
    Expert & 88.11 & 89.14 & 93.95 \\
    Model & 92.73 & 92.79 & 92.81 \\
    \bottomrule
    \end{tabular}}%
  \label{tab:hatespeech}%
\end{table}%

\subsubsection{Results on CIFAR10H and GalaxyZoo} \label{Sec:human_label}

\begin{table}[b]
  \caption{Results on CIFAR10H and GalaxyZoo with real human labels. * denote results of other methods reproduced by us.}
  \centering{\small
    \begin{tabular}{ccccc}
    \toprule
          & \multicolumn{2}{c}{CIFAR10H} & \multicolumn{2}{c}{GalaxyZoo} \\
          & Mean  & S.E.  & Mean  & S.E. \\
    \midrule
    Confidence & 95.09  & 0.40  &  81.83* & 0.20 \\
    Consistent (Impute) & 96.03  & 0.21  & 83.40* & 0.19 \\
    Consistent (2-step) & \underline{96.29}  & 0.25  & 81.97* & 0.23 \\
    DiffTriage & -  & -  & \underline{83.80}  & 0.35 \\ 
    Equal Weights & 94.70  & 0.00  & 80.19  & 0.48  \\
    Pooled Weights & 96.20  & 0.31  & 83.13  & 0.12  \\
    Flexible Weights & \textbf{96.53}  & 0.32  & \textbf{84.46}  & 0.18  \\
    \bottomrule
    \end{tabular}%
  \label{tab:human_label}}%
\end{table}%

CIFAR10H and Galaxyzoo are image classification datasets with real human labels. For CIFAR-10H, we split it with 0.7:0.15:0.15 for training set, validation set and test set. For each example, we random sample an expert's label in each epoch. For Galaxyzoo, we apply the same pre-processing method as in \citep{okati2021differentiable} with 0.6:0.2:0.2 train-val-test split ratio, and perform binary classification with the weighted combination of both human score vectors and model outputs. We compare the proposed methods with \citep{raghu2019algorithmic}, \citep{Hussein2020Consistent} and \citep{okati2021differentiable}. A pretrained WideResNet-28 with the same setting as in \citep{Hussein2020Consistent} is applied for finetuning the AI model as well as the feature extractor for both CIFAR10H and GalaxyZoo datasets. For our method, we apply three settings: The first is using equal weights for both human and AI model (denoted by ``Equal Wights''). The second is using trained global weights, which means we update the weights for every example in each iteration, and then take average of them to get shared global weights for human and AI. The global weights are applied for calculating the combination scores (denoted as ``Pooled weights''). The third is flexible weights, which is using instance aware weights for calculating the combination scores (denoted as ``flexible weights''). We apply an Adam optimizer with learning rate 1e-4 for the weights generation model, and 1e-5 for pretrained AI model and feature extractors. The effective batch size is set to be 128. The checkpoints with the minimum validation loss in 50 epochs of training are applied for testing. 

The results are given in Table~\ref{tab:human_label}. We report both the average accuracy of 5 trials with different seeds and the standard error of the average. As not all related papers report the numerical results on both two dataset, we reproduce the rest by ourselves. We observe that for both two datasets, the flexible weights setting achieves the highest classification accuracy. Our method with flexible weights setting significantly outperforms the method in \citep{okati2021differentiable} on Galaxyzoo dataset, and the method in \citep{raghu2019algorithmic} on CIFAR-10H. It achieves the higher average performance than \citep{Hussein2020Consistent} on CIFAR-10H although the difference is less significant. Our reproduced accuracy of DiffTriage on CIFAR10H is lower than model-only accuracy ($<$93\%), which may due to the lack of tuning or the ineffectiveness of this approach on the CIFAR-10H dataset (more categories than Galaxyzoo and hatespeech).

\begin{figure}[t]
\begin{center}
 \includegraphics[width=1.0\linewidth]{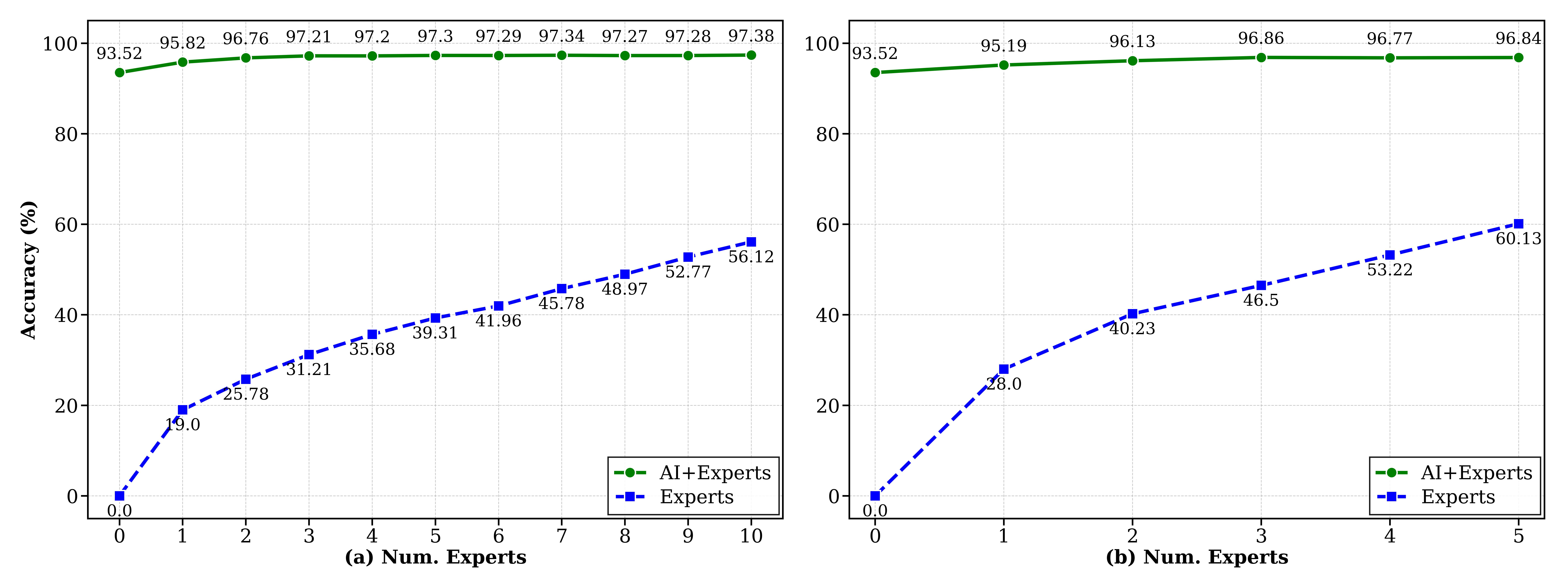}
 \caption{Diagram of accuracy under collaborative decision by AI models and multiple human experts.} 
 \label{Fig:mul_exp}
\end{center}
\end{figure}

\subsection{Collaborative decision by one model and multiple human experts.}

To investigate the effectiveness of the proposed architecture for the scenario with multiple human experts, we assume that we have multiple human experts with different expertise sets with equal size, for which each expert can have only 1 or 2 expertise categories. We consider the collaborative decision making by one model and these human experts. Also, we consider the case of solely using these experts for collaborative decision making without collaboration with AI model. The experiments are done on CIFAR10. We adopt a pretrained ResNet-18 model as the classifier as well as the image representation extractor. A 6 layer transformer encoder with 8 heads and a embedding dimension of 512 is applied. An Adam optimizer with a learning rate of 1e-5, and a batch size of 64 are applied. The result of the case that each expert has only 1 expertise category is shown in Figure~\ref{Fig:mul_exp}(a), where $s$ is the size of experts group. We observe that when the number of experts is small, the performance of both AI+experts system and experts-only system increase linearly. However, when the number of experts becomes large, the increasing trends diminishes slowly. This indicates that we need to apply heavy hyper-parameter tuning or use a more powerful weight generation model for handling the collaboration with a large number of experts.
In the second experiment of multiple expert collaborative decision making. we assume that each expert has an expertise set with size 2, and the 5 experts have expertise sets $\{1,2\}$, $\{3,4\}$, $\{5,6\}$, $\{7,8\}$, $\{9,10\}$, respectively. The result of this case that is shown in Figure~\ref{Fig:mul_exp}(b), where We observe that similar trend as in Figure~\ref{Fig:mul_exp}(a). The propose method provide a significant improvement of the system accuracy as number of expert increases, but there is a diminishing marginal utility effect when the expert number is large.



\section{Analysis}


\label{Sec:7.2}
\begin{figure}[t]
\begin{center}
 \includegraphics[width=1.0\linewidth]{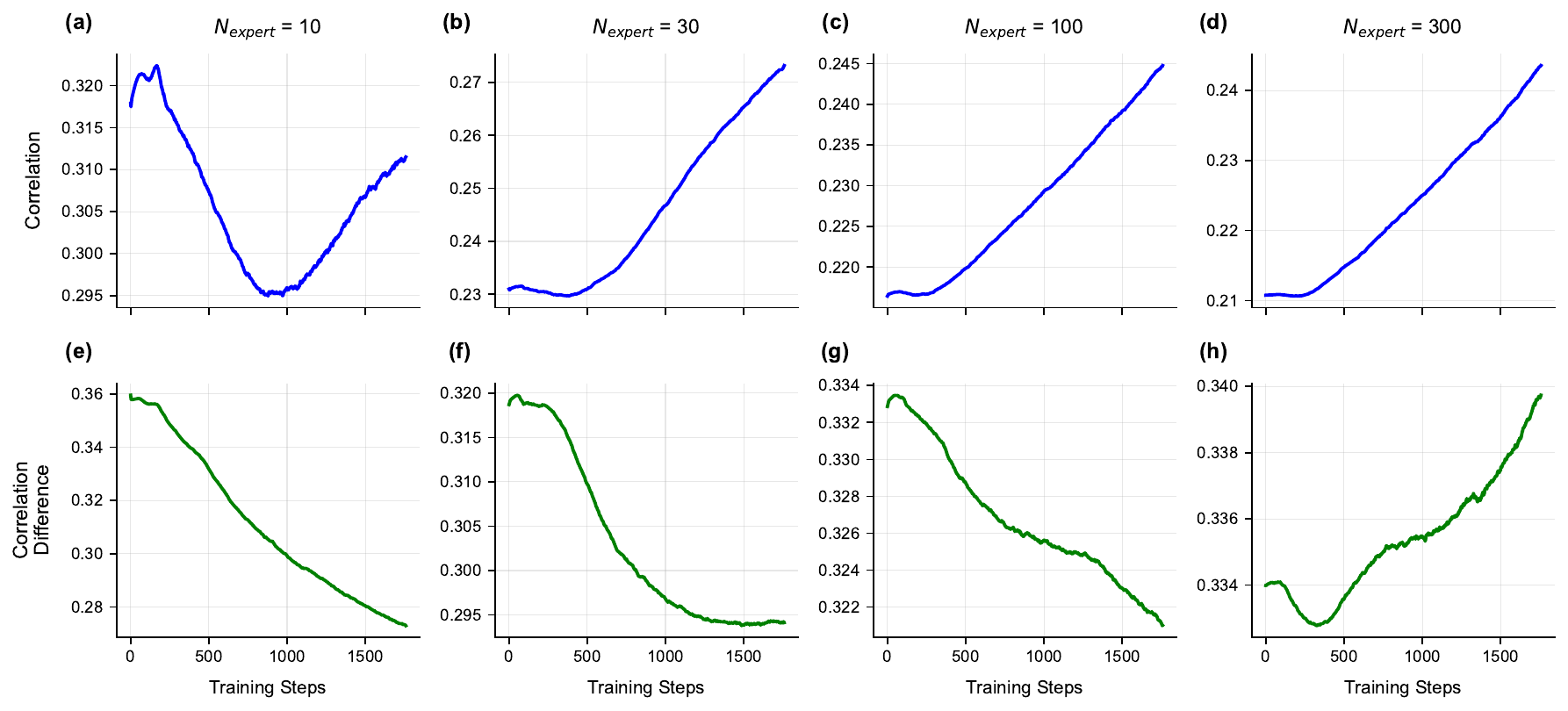}
 \caption{Diagram of the learning curves of capability vectors including: (a) The average of absolute correlation between capability vectors and (b) The average of absolute difference between the correlation of capability vectors and correlation of corresponding binary vectors of expertise set.} 
 \label{Fig:corr}
\end{center}
\end{figure}

\subsection{Learning behavior of capability vectors with different vocabulary sizes.} 
For the experiments of random sampled decision makers with different vocabularies (see Section~\ref{Sec:6.3.2}), we investigate the change of average absolute correlations between capability vectors with respect to the learning steps, which is shown in Figure~\ref{Fig:corr}(a). We set different vocabulary sizes in $\{10, 30, 100, 300\}$, and apply a single linear mapping from one-hot decision maker representations to the capability embeddings with size 16, which have the same size with the hidden embedding. 
We observe that when the experts' vocabulary is relatively large, there is an increasing trend of the correlation between capability vectors as the training step gets larger. Our explanation is that as there are in total 1024 different capability settings, when the number of decision makers becomes larger, the change of modeling similar capability also becomes larger, resulting in a larger average correlation value. 
Furthermore, we investigate whether the correlation between the learned capability vectors converges towards the correlation between the underlying binary vectors representing the capabilities. To quantify this, we compute the absolute differences between the two correlation matrices and track the average deviation over the course of training. Specifically, at the $t$-th training step, this metric is defined as:
\begin{equation}
D_{corr,t} = \frac{1}{M^2}\sum_{i=1}^M \sum_{j=1}^M |\rho^{\text{cap}}_{ij, t}-\rho^{\text{bin}}_{ij, t}|,
\end{equation}
where $M$ is the vocabulary size of decision makers, $\rho^{\text{cap}}_{ij, t}$ and $\rho^{\text{bin}}_{ij, t}$ are the entries of the correlation matrix between the learned capability vectors in the vocabulary and the correlation matrix between the binary vectors of underlying pre-defined expertise sets of decision makers, respectively, and step $t$. The results are shown in Figure~\ref{Fig:corr}(b). Notably, for vocabulary sizes of 10, 30, and 100, we observe a clear trend of decreasing average absolute differences between the two correlation matrices. 
As for the case of 300 decision makers in the vocabulary, $D_{corr,t}$ slightly increases from 0.334 to 0.339, which may due to the learning of too many capability vectors simultaneously makes it hard to capture the correlations.


\section{Discussion}

\subsection{Comparing with related works}
The proposed architecture adopts a strategy of merging the output vectors from both AI models and human experts, rather than merely deciding whether to defer the task to a human. By applying a dynamical weighted combination of the output vectors from both human and AI agents, our method outperforms the major learning to defer methods. In contrast to the work by \citep{kerrigan2021combining}, which also focuses on output vector combination, our method incorporates a representation learning mechanism instead of relying on a naive Bayesian estimator based on a confusion matrix. This allows our approach to seamlessly accommodate multiple experts in an end-to-end manner.

Our method also facilitates human-AI collaborative decision-making involving multiple AI and human agents. This is achieved through the learning of capability vectors in a uniform dimensional space. Related studies on human-AI collaborative decision-making with multiple experts typically employ different datasets or base models in their experiments. For instance, \citep{mao2024two} utilize ResNet-4 as the predictive AI model, while employing ResNet-10, ResNet-16, and ResNet-28 with increasing capacities as the human experts, focusing on the CIFAR-10 and SVHN image classification datasets. Additionally, \citet{Hemmer2022Forming} investigate three distinct tasks, including Hate Speech detection, CIFAR-100 classification, and the NIH Dataset. As different papers apply different experimental settings, it is hard to make a complete comparison. Current experimental results convince us the advantage of our method can be extend to other settings.

\subsection{Potential application scenarios}

In addition to traditional human-machine collaborative decision-making scenarios, the proposed method can be applied to a broader range of situations, specifically including collaborative labeling, decision participant selection, and large-scale multi-task training.

\textbf{Collaborative Annotation}: Crowdsourcing platforms typically employ a large number of temporary annotators. For each annotation task, annotators are randomly assigned to perform the labeling. However, due to variations in the skills and areas of expertise among different annotators, the quality of the annotations may be compromised. By employing a collaborative decision-making model based on capability vectors, we can learn the capability vectors of annotators from historical annotation data. This model not only facilitates collaboration among multiple annotators based on their capability vectors but also allows for the integration of annotations from both the annotators and auxiliary models. This approach can enhance the quality of annotations and improve overall annotation efficiency.
 
\textbf{Decision maker selection}: Based on the assumptions of our method, the performance of collaborative decision making can be determined by the capability vectors of decision makers, as well as the task embedding. Therefore, given the task instance, the performance can be determined by the set selection of decision makers. Further, given the task instance and part of the decision makers, the performance can be determined by the rest decision makers. We can build a predictive model for predicting the final performance of collaborative decision making based on decision makers' capability vectors and the task embedding. This can be used for selecting part or all decision makers for each particular task instance. One example is assigning editor/reviewers for research article reviewing, where each editor/reviewer can be considered as a decision maker with an associated capability vector.

\textbf{Large-scale multitask training}: Our method can be applied for classification tasks, regression tasks and generation tasks. For comparing the existing methods we focus on classification task in our experiments. For regression task, capability vectors of decision makers can determine the weights for the predicted values from all the decision makers. As to generation task, we can apply a weighted combination of scores from multiple decision makers on each of multiple generated results (e.g. generated by human writers and/or a set of LLM agents), and use it to determine which generated result to be selected. The decision makers' capability vectors can be applied for a variety of related tasks, which is comparable to the training of token embeddings on multiple language processing tasks with transformer-based models. The trained capability vectors can potentially be generalized to a larger set of tasks.  

\section{Conclusion}

In this study, we present a human-AI collaborative decision-making architecture based on capability vectors, applicable to both generative and discriminative decision tasks. By utilizing capability vectors, we can uniformly model the decision intelligence of human experts and AI models across the same set of dimensions. Additionally, we developed a learning-free global weighted merging method for collaborative decision-making involving one human expert and one AI model specifically for classification tasks, which can be applied as a baseline for check the effectiveness of instance-aware weighted merging methods. 
Our experiments, conducted on image classification and sentiment analysis tasks, demonstrate that the proposed method outperforms existing approaches across various human-AI collaborative scenarios. Experiments are performed using both simulated experts on CIFAR10/CIFAR100/Hatespeech and real experts on CIFAR-10H and GalaxyZoo, which demonstrate the advantage of proposed method over existing state-of-the-art models. We also show that for a widely used expert model in classification tasks, our method is more robust to the change of non-expertise capability level. In addition, the proposed method can be applied for multiple expert collaborative decision with or without AI decision makers. 

In future works, empirical studies involving real human experts can be done to further investigate the effectiveness of our architecture when applied in real world. More complex human decision modelings with confidence scores can also be explored.
Moreover, our method can be extended to collaborative decision-making for evaluating generated results, which holds significant potential for applications in managerial decision-making and textual answer evaluation. We believe there remains considerable opportunity for further exploration of the proposed architecture, which is left for the future studies.

\bibliography{Reference}

\end{document}